\documentclass[11pt]{article}

\usepackage{amsmath,amssymb,epsfig,cite,color,verbatim,array,bm,amsfonts,enumerate,enumitem,mathabx}
\usepackage{graphics,float,multirow}
\usepackage[colorlinks=true,citecolor=ForestGreen,urlcolor=black,linkcolor=blue]{hyperref}
\usepackage{blindtext}
\usepackage{microtype}
\usepackage[latin9]{inputenc}
\usepackage{authblk,booktabs,hhline}
\usepackage[table,dvipsnames]{xcolor}

\definecolor{light-gray}{gray}{0.9}

\newcommand*{\approxident}{
\mathrel{\vcenter{\offinterlineskip
\hbox{$\sim$}\vskip-.35ex\hbox{$\sim$}\vskip-.35ex\hbox{$\sim$}}}}
\numberwithin{equation}{section}
\allowdisplaybreaks
\title{(Extended) Proca-Nuevo under the two-dimensional loupe }

\author[1,2]{Ver\'onica Errasti D\'iez \footnote{vero.erdi@origins-cluster.de}}

\affil[1]{ {\small {\it Universit\"ats-Sternwarte, Fakult\"at f\"ur Physik,
Ludwig-Maximilians-Universit\"at M\"unchen, Scheinerstra\ss e 1,
D-81679 M\"unchen, Germany}}}

\affil[1]{ {\small {\it Excellence Cluster ORIGINS, Boltzmannstra\ss e 2,
D-85748 Garching bei M\"unchen, Germany}}}

\date{}

\begin{document}

\maketitle

\begin{abstract}
Recently, two new families of non-linear massive electrodynamics have been proposed: Proca-Nuevo and Extended Proca-Nuevo.
We explicitly show that both families are irremediably ghostful in two dimensions. Our calculations indicate the need to revisit the classical consistency of (Extended) Proca-Nuevo in higher dimensions before these settings can be regarded as ghostfree.
\end{abstract}

\section{Introduction}
\label{sec:intro}

Ever since Maxwell unified the electric and magnetic forces in the mid-$19^{\textrm{th}}$ century~\cite{Maxwell},
his theory has played a pivotal role in physics.
In particular, it is the conventional framework for
electromagnetism.
Nonetheless, other frameworks do exist.
In the following, we provide the fleeting chronology of some of them.

We restrict attention to theories that admit a local Lagrangian formulation in terms of a real Abelian vector field $A_\mu$
and that are Lorentz-invariant and first-order.
By first-order we mean electrodynamics theories whose Lagrangian does {\it not} depend on second- or higher-order derivatives.
Influential theories beyond our scope encompass Podolsky~\cite{Pod1,Pod2} (higher-order), Yang-Mills~\cite{Yang:1954ek} (non-Abelian),
BF~\cite{Schwarz:1978cn} (extra field content), the vector sector of the minimal Standard Model Extension (SME)~\cite{Kostelecky:2002hh}
(Lorentz-violating) and constructions by Mashhoon~\cite{Mashhoon:2004rz,Mashhoon:2005ft,Mashhoon:2007sf} (non-local).
We also dismiss Chern-Simons~\cite{Chern:1974ft}, on the account that it can only be defined in an odd number of spacetime dimensions.
For a shrewd presentation of the exceptionality of Podolsky and Chern-Simons, see~\cite{Fonseca:2010av}.

The reader should not be dispirited by our omissions:
the kinds of alternatives to Maxwell electrodynamics that we do consider are relevant and plentiful.
They date back 
to the early $20^{\textrm{th}}$ century.
In the wake of Quantum Mechanics, initial attempts at quantizing Maxwell electrodynamics 
met with an acute problem: in the presence of electromagnetic sources, the self-energy of the electron seemed to diverge.
So as to circumvent the hurdle, modifications to Maxwell's theory
proliferated~\cite{Bovy}. 
As is well-known, the conundrum was ultimately resolved in agreement with Maxwell's original proposal~\cite{Johnson}.
However, the misled efforts were not in vain:
despite their venturesome genesis, modified electrodynamics theories
are widely employed and new systems are still introduced, with diverse applications in mind.

Two types of generalizations
of Maxwell's theory must be distinguished.
On the one hand, the massive electrodynamics theory
put forward by Proca~\cite{Proca1,Proca2},
which quickly became and remains bedrock to optics~\cite{Bloem,Photonics}.
On the other hand, a populous family of non-linear (massless) electrodynamics theories,
all of which belong in
the {\it Non-Linear Electrodynamics} (NLE) class~\cite{Pleb}. 
Prominent examples include the long-lived Born-Infeld~\cite{Born:1934gh} and the recent ModMax~\cite{Bandos:2020jsw,Kosyakov:2020wxv}.
The reader will find in~\cite{Sorokin:2021tge} a succinct and compelling review of NLE, including its multi-disciplinary utility.

It is natural to supplement Proca's theory
with self-interaction terms that depend on the vector field $A_\mu$ or also on its field strength $F_{\mu\nu}$\footnote{
Unfortunately, the ubiquity of these self-interactions has prevented us from identifying their first appearance.}.
For instance, consider quartic vertices for the $Z$ boson after electroweak symmetry breaking~\cite{Eboli:2016kko},
Beyond the Standard Model physics~\cite{Freitas:2021cfi}
and boson stars~\cite{Aoki:2022woy}.
We collectively refer to such non-linear extensions of Proca electrodynamics as {\it Self-interacting Proca}.

Contrastingly, the inclusion of derivative self-interactions that are {\it not} confined to the field strength $F_{\mu\nu}$ is non-trivial.
The idea took
almost 80 years to materialize, in the shape of the {\it Generalized Proca} class~\cite{Tasinato:2014eka,Heisenberg:2014rta}\footnote{A subset of Generalized Proca was independently rediscovered in~\cite{Eboli:2016kko}.}.
The motivation came from cosmology:
in the presence of gravity,
Generalized Proca settings are natural candidates to explain the late-time
accelerated expansion of our universe.
The original proposal
underwent some controversy~\cite{Allys:2015sht}, but the complete
Lagrangian was obtained shortly after~\cite{BeltranJimenez:2016rff}.
Generalized Proca contains Self-interacting Proca as a subset:
$\mathcal{L}_2$ in equation (12) of~\cite{BeltranJimenez:2016rff}.

Up to this point, the plethora of considered electrodynamics theories admits the classification in table \ref{table:before}.
In terms of their constraint structure, 
massless (massive) electrodynamics are associated with two first (second) class constraints. The change in the constraints' class happens because all massive electrodynamics explicitly break the (manifest) $U(1)$ gauge symmetry of the massless sector.

\begin{table}[!ht]  
\renewcommand*{\arraystretch}{1.9}
\centering
\begin{tabular}{|>{\centering\arraybackslash}m{3cm}|>{\centering\arraybackslash}m{3cm}|>{\centering\arraybackslash}m{9cm}|}
\hhline{~|-|-|}
\multicolumn{1}{c|}{} & \cellcolor{light-gray}  {\bf Linear} &  \cellcolor{light-gray}  {\bf Non-linear} \\ \hline
\cellcolor{light-gray}  {\bf Massless} & Maxwell   & Non-linear electrodynamics (NLE)   \\ \hline
{\cellcolor{light-gray}{\bf Massive}} &Proca  & Generalized Proca   \\ \hline
\end{tabular}
\caption{Classification of electrodynamics theories within scope prior to~\cite{deRham:2020yet,deRham:2021efp}.}
\label{table:before}
\end{table}

Latterly, a physically inequivalent class of non-linear completions to Proca electrodynamics, dubbed {\it Proca-Nuevo}, has appeared~\cite{deRham:2020yet}. It has further been suggested that Proca-Nuevo and specific interactions of the Generalized Proca class can be combined into {\it Extended Proca-Nuevo}~\cite{deRham:2021efp}.
As with Generalized Proca, cosmology triggered and fuels the ongoing interest in (Extended) Proca-Nuevo~\cite{deRham:2020yet,deRham:2021efp,deRham:2021yhr,Pozsgay:2022xpc,deRham:2022sdl}.

Accordingly, the up-to-date vast landscape of electrodynamics theories within scope can be summarized in table \ref{table:after}.
For additional details on the non-linear massive sector, see figure 1 in~\cite{deRham:2021efp}. The massive electrodynamics character of (Extended) Proca-Nuevo requires that both classes are associated with two second class constraints.

\begin{table}[!ht]  
\renewcommand*{\arraystretch}{1.75}
\centering
\begin{tabular}{|>{\centering\arraybackslash}m{3cm}|>{\centering\arraybackslash}m{3cm}|>{\centering\arraybackslash}m{4.5cm} |>{\centering\arraybackslash}m{4.5cm}|}
\hhline{~|-|-|-|}
\multicolumn{1}{c|}{} & \cellcolor{light-gray}  {\bf Linear} & \multicolumn{2}{c|}{\cellcolor{light-gray}  {\bf Non-linear}} \\ \hline
\cellcolor{light-gray}  {\bf Massless} & Maxwell   & \multicolumn{2}{c|}{Non-linear electrodynamics (NLE)}   \\ \hline
\multirow{2}{*}{\cellcolor{light-gray}{\bf Massive}}&\multirow{2}{*}{Proca}&
Generalized Proca  & \multirow{2}{*}{Extended Proca-Nuevo}\\
\cline{3-3}
\multirow{-2}{*}{\cellcolor{light-gray}{\bf Massive}}&\multirow{2}{*}{}&Proca-Nuevo  &\multirow{2}{*}{}\\
 \hline
\end{tabular}
\caption{Current classification of electrodynamics theories within scope.}
\label{table:after}
\end{table}

This manuscript is devoted to the comprehensive study of the classical consistency of the latest two families of electrodynamics theories: (Extended) Proca-Nuevo. For algebraic simplicity, we work in two spacetime dimensions.
In this toy-scenario, we find that both families are never related to two second class constraints, which forces their rejection and a rewind in the understanding of massive electrodynamics: from table \ref{table:after} back to table \ref{table:before}. We affirm the need for further studies
before a choice between tables \ref{table:before} and \ref{table:after}
is made for higher dimensions.

The text is organized as follows.
In section \ref{sec:review}, we review the Proca-Nuevo class of electrodynamics theories in full generality.
From section \ref{sec:2dcase} onward, we focus on two spacetime dimensions.
We begin by conveniently reformulating the Lagrangian in section \ref{sec:2dcase}.
This provides the starting point for the explicit
constraint analysis in section \ref{sec:cons},
where we show that the setup is generically ghostful.
In section \ref{sec:limit}, we demonstrate that there does not exist a ghostfree limit to Proca-Nuevo.
In section \ref{sec:rigid}, we prove that Proca-Nuevo cannot be extended
via the addition of terms from other theories in table \ref{table:before} in a ghostfree manner. 
In particular, we thus rule out Extended Proca-Nuevo.
We draw our conclusions in the final section \ref{sec:concl}.
Supplementary mathematical details can be found in the appendix \ref{app:math}.

\vspace*{0.5cm}

\noindent
\textbf{Conventions.}\\
We work in Minkowski spacetime $\mathcal{M}$.
We begin by presenting the setup for arbitrary but finite dimension $d\geq 2$.
For this part, spacetime indices are denoted by the Greek letters $(\mu,\nu,\ldots)$ and raised/lowered
with the mostly positive metric $\eta_{\mu\nu}=\textrm{diag}(-1,1,1,\ldots)$ and its inverse $\eta^{\mu\nu}$.
We employ the short-hand $\displaystyle\partial_\mu:=\frac{\partial}{\partial x^\mu}$,
where $x^\mu$ are the spacetime local coordinates.
Henceforth, we set $d=2$.
We make use of dots to indicate time derivatives:
$\displaystyle\dot{f}:=\partial_0f$ and $\displaystyle\ddot{f}:=\partial_0^2f$, for any local function $f$.
The Latin letter $a$ designates physically inequivalent branches of the two-dimensional setup.
Einstein summation applies for spacetime indices but {\it not} for the branch label $a$.
When needed, sum over $a$ is explicitly stated.
Natural units are considered.

\section{Review of Proca-Nuevo}
\label{sec:review}

In this section, we provide a brief, technical presentation of the {\it Proca-Nuevo} class of electrodynamics theories~\cite{deRham:2020yet}.
The forthcoming review helps us set our notation and contextualize our results.

We start by introducing a few auxiliary tensors.
On $\mathcal{M}$, define the symmetric $(0,2)$ tensor
\begin{align}
\label{eq:deff}
f_{\mu\nu}:=\eta_{\mu\nu}+\frac{1}{\Lambda^2}\left(\partial_{\mu}A_{\nu}+\partial_\nu A_\mu\right)+\frac{1}{\Lambda^4}\partial_\mu A_\rho \partial_\nu A^\rho,
\end{align}
where $\Lambda$ is a constant of length dimension $(-1)$.
Using the above, the $(1,1)$ tensor $\chi^\mu{}_\nu$ is defined via its square:
\begin{align}
\label{eq:defchi}
\left(\chi^2\right)^\mu{}_\nu\equiv \chi^\mu{}_\rho \chi^\rho{}_\nu:= \eta^{\eta\rho}f_{\rho\nu}.
\end{align}
The latter definition (\ref{eq:defchi}) motivates the symbolic formula
\begin{align}
\label{eq:symchi}
\chi^\mu{}_\nu=\left(\sqrt{\eta^{-1}f}\right)^\mu{}_\nu
\end{align}
that is nuclear to Proca-Nuevo 
and that serves to further define another $(1,1)$ tensor
\begin{align}
\label{eq:defcalK}
\mathcal{K}^{\mu}{}_\nu:=\chi^\mu{}_\nu-\delta^\mu{}_\nu.
\end{align}
We note that, in postulating Proca-Nuevo, $\chi^\mu{}_\nu$ is assumed to be invertible.
Namely, it is (implicitly) taken for granted that there exists $\left(\chi^{-1}\right)^\mu{}_\nu$ such that 
\begin{align}
\label{eq:chiinv}
\left(\chi^{-1}\right)^\mu{}_\rho \chi^\rho{}_\nu=\delta^\mu{}_\nu=\chi^\mu{}_\rho\left(\chi^{-1}\right)^\rho{}_\nu.
\end{align}

In the Lagrangian formulation, Proca-Nuevo
can be succinctly expressed as
\begin{align}
\label{eq:PNaction}
S_{\textrm{PN}}=\int_{\mathcal{M}}d^dx\,\,
\Lambda^d \mathcal{L}_{\textrm{PN}}, \qquad
\mathcal{L}_{\textrm{PN}}= 
\sum_{n=0}^d \alpha_n \mathcal{L}_n\left(\mathcal{K}\right),
\end{align}
where the $\alpha_n$'s are arbitrary smooth functions of the (suitably normalized) vector field squared,
\begin{align}
\label{eq:alphas}
\alpha_n=\alpha_n\left(\frac{A_\mu A^\mu}{\Lambda^2}\right)\equiv\alpha_n\left(\frac{A^2}{\Lambda^2}\right),
\end{align}
and where $\mathcal{L}_n\left(\mathcal{K}\right)$ stands for the univariate
elementary symmetric polynomial (ESP) of degree $n$,
\begin{align}
\label{eq:mathlps}
\mathcal{L}_n=e_n\left(\mathcal{K}^\mu{}_\nu\right).
\end{align}
Such ESPs depend exclusively
on the trace of various powers of $\mathcal{K}^\mu{}_\nu$,
which we shall denote $[\mathcal{K}^n]\equiv \left(\mathcal{K}^n\right)^\mu{}_\mu\equiv \mathcal{K}^\mu{}_{\nu_1}\chi^{\nu_1}{}_{\nu_2}\ldots \mathcal{K}^{\nu_{n-1}}{}_\mu$.
The explicit form of the three lowest degree ESPs
will come in handy shortly:
\begin{align}
\label{eq:elempols}
e_0\left(\mathcal{K}^\mu{}_\nu\right)=1, \qquad e_1\left(\mathcal{K}^\mu{}_\nu\right)=[\mathcal{K}],
\qquad e_2\left(\mathcal{K}^\mu{}_\nu\right)=\frac{1}{2}\left([\mathcal{K}]^2-[\mathcal{K}^2]\right).
\end{align}
The reader can find a clear introduction to ESPs and their chief properties in~\cite{ESPs}.
We remark that ESPs are the building blocks of other theories, including
non-linearly massive gravity theories~\cite{Hassan:2011vm} and Generalized Proca (\ref{eq:ESpSEPN}).
Overall, the Proca-Nuevo action (\ref{eq:PNaction})
is a functional of $A^2$ and $[\mathcal{K}^n]$
or $[\chi^n]$.
We work in terms of the latter form
$\mathcal{L}_{\textrm{PN}}=\mathcal{L}_{\textrm{PN}}\left[A^2,[\chi^n]\right]$.

We conclude our lightning presentation of Proca-Nuevo with two important remarks.
First, $f_{\mu\nu}$ in (\ref{eq:deff})
is a manifestly Poincar\'e-covariant object.
Since this is the building block of the action (\ref{eq:PNaction}), theories within the
Proca-Nuevo class are Lorentz-invariant~\cite{deRham:2020yet,deRham:2021yhr,deRham:2021efp}. 
Second, Proca-Nuevo is {\it not} a class of gauge theories.
In other words, there is no gauge transformation under which
(\ref{eq:PNaction}) remains unchanged.
In agreement with our observation in the introduction section \ref{sec:intro}, the $U(1)$ gauge symmetry inherent to massless electrodynamics
is explicitly broken.
Both considerations will play a role later.

\section{Two-dimensional Proca-Nuevo}
\label{sec:2dcase}

Given the general framework in the previous section \ref{sec:review},
the Lagrangian density for the Proca-Nuevo class in two spacetime dimensions is
\begin{align}
\label{eq:2dLag}
\mathcal{L}_{\textrm{PN}_2}=
\alpha_0+\alpha_1[\chi]+\alpha_2\left([\chi]^2-[\chi^2]\right),
\end{align}
where, for simplicity and without loss of generality,
we have redefined the $\alpha$'s as
\begin{align}
\label{eq:newalphasdef}
\alpha_0^{(\textrm{new})}:= \alpha_0-2\alpha_1, \qquad
\alpha_1^{(\textrm{new})}:= \alpha_1-\alpha_2, \qquad
\alpha_2^{(\textrm{new})}:= \alpha_2/2
\end{align}
and dropped the ``new'' labels.
It follows from the definitions (\ref{eq:deff}) and (\ref{eq:defchi}) that $\left(\chi^2\right)^\mu{}_\nu$ is known explicitly:
\begin{align}
\label{eq:chi2}
\begin{array}{lllll}
\mathbf{A}\equiv
\left(\chi^2\right)^0{}_0=
\displaystyle
1+\frac{\dot{A}_0}{\Lambda^2}
\left(\frac{\dot{A}_0}{\Lambda^2}-2\right)
-\frac{\dot{A}_1^2}{\Lambda^4}, \qquad\quad
\mathbf{B}\equiv 
\left(\chi^2\right)^1{}_1=
\displaystyle
1+\frac{\partial_1A_1}{\Lambda^2}
\left(\frac{\partial_1A_1}{\Lambda^2}+2\right)
-\frac{\left(\partial_1 A_0\right)^2}{\Lambda^4},
\vspace*{0.3cm}\\
\hspace*{2.5cm}\mathbf{C}\equiv 
\left(\chi^2\right)^0{}_1=
\displaystyle
-\left(\chi^2\right)^1{}_0=
-\frac{\dot{A}_1}{\Lambda^2}
\left(\frac{\partial_1 A_1}{\Lambda^2}+1\right)
+\frac{\partial_1A_0}{\Lambda^2}
\left(\frac{\dot{A}_0}{\Lambda^2}-1\right).
\end{array}
\end{align}
Therefore, $[\chi^2]$ in (\ref{eq:2dLag}) is known explicitly: $[\chi^2]=\mathbf{A}+\mathbf{B}$.
In contrast, the symbolic formula (\ref{eq:symchi})
does {\it not} allow for the straightforward inference
of neither $\chi^\mu{}_\nu$ nor its trace $[\chi]$
appearing in (\ref{eq:2dLag}).
By means of (\ref{eq:symchi}), we can only conclude that $\chi^0{}_1=-\chi^1{}_0$.

To find $[\chi]$, we solve the matrix equation
\begin{align}
\label{eq:matrix2}
\chi^\mu{}_\nu\chi^\nu{}_\rho=\left(\chi^2\right)^\mu{}_\rho,
\end{align}
where the left hand side is unknown and the right hand side is given by (\ref{eq:chi2}).
This is a system of three functionally independent 
quadratic equations in three variables,
\begin{align}
\label{eq:quartic}
\left(\chi^0{}_0\right)^2-\left(\chi^0{}_1\right)^2
-\mathbf{A}=0,  \qquad
\left(\chi^1{}_1\right)^2-\left(\chi^0{}_1\right)^2
-\mathbf{B}=0, \qquad 
\chi^0{}_1\left(\chi^0{}_0+\chi^1{}_1\right)-\mathbf{C}=0,
\end{align}
and as such it can be solved for arbitrary real values of $\left(\mathbf{A},\mathbf{B},\mathbf{C}\right)$.
There exist four solutions\footnote{\label{foo:othersols} Other solutions to (\ref{eq:quartic}) do exist, but they require that we impose non-trivial relations among $\left(\mathbf{A},\mathbf{B},\mathbf{C}\right)$. This means postulating stringent constraints on the vector field $A_\mu$, which would set our work apart from the literature. We dismiss such cases.}; divided into two, a priori physically inequivalent branches
$\pm\chi^{(\pm)}$, with 
\begin{align}
\label{eq:chiexpl}
\left(\chi^{(\pm)}\right)^0{}_0=\frac{\mathbf{A}\pm\mathbb{D}}{\mathbb{E}}, \qquad 
\left(\chi^{(\pm)}\right)^0{}_1=\frac{\mathbf{C}}{\mathbb{E}},
\qquad
\left(\chi^{(\pm)}\right)^1{}_1=\frac{\mathbf{B}\pm\mathbb{D}}{\mathbb{E}}, 
\end{align}
where we have introduced
\begin{align}
\label{eq:defD}
\mathbb{D}:=\sqrt{\mathbf{A}\mathbf{B}+\mathbf{C}^2},
\qquad
\mathbb{E}:=\sqrt{\mathbf{A}+\mathbf{B}\pm 2 \mathbb{D}}.
\end{align}
In agreement with equation (\ref{eq:chiinv}) and footnote \ref{foo:othersols}, the solutions (\ref{eq:chiexpl}) entail
\begin{align}
\label{eq:req}
\mathbf{C}\neq 0, \qquad 
\mathbb{E}\neq 0.
\end{align}
It readily follows that $[\chi^{(\pm)}]=\mathbb{E}$.
It is worth noting that, in terms of the vector field $A_{\mu}$, $\left(\mathbf{A}\mathbf{B}+\mathbf{C}^2\right)$ is a perfect square.
Henceforth and without loss of generality, we drop the overall sign duplicity
within each branch --- it can be absorbed by a suitable redefinition of
$\alpha_1$ in (\ref{eq:2dLag}).
We also relabel the two branches of the setup: hereafter, $a=1$ refers to the minus branch,
while $a=2$ denotes the plus branch.

On the whole and after an irrelevant rescaling of $\alpha_2$, the Lagrangian density (\ref{eq:2dLag}) takes the form
\begin{align}
\label{eq:2dLagexpl}
\mathcal{L}_{\textrm{PN}_2}=\alpha_0
+\alpha_1\mathbb{E}+\alpha_2\mathbb{D},
\end{align}
which encodes two a priori distinct physics
for the vector field $A_\mu$.
For our subsequent purposes, it is convenient to make the two branches apparent and 
rewrite the above as
\begin{align}
\label{eq:2dLagfin}
\mathcal{L}_{\textrm{PN}_2}^{(a)}=
\alpha_0
+\alpha_1{N}_a
+\alpha_2\sum_{a=1}^2(-1)^a{N}_a^2,
\end{align}
where we have defined
\begin{align}
\label{eq:defxys}
\begin{array}{llll}
&\displaystyle N_a:=\sqrt{x_a^2-y_a^2}, &\displaystyle\qquad 
x_1:=\frac{1}{\Lambda^2}\left(\dot{A}_0+\partial_1 A_1\right), \vspace*{0.3cm} \\
&\displaystyle x_2:=\frac{1}{\Lambda^2}\left(-\dot{A}_0+\partial_1 A_1\right)+2,
&\displaystyle\qquad 
y_a:=\frac{1}{\Lambda^2}\left[\dot{A}_1-(-1)^a\partial_1 A_0\right].
\end{array}
\end{align}
The explicit example in~\cite{deRham:2020yet}
follows from setting
\begin{align}
\label{eq:exchoices}
\alpha_0=4-\frac{1}{2}m^2{A_\mu A^\mu}, \qquad
\alpha_1=-2, \qquad
\alpha_2=0,
\end{align}
with $m\in\mathbb{R}^+$,
within the (plus) branch  $\mathcal{L}_{\textrm{PN}_2}^{(2)}$.

For later convenience, we also define 
\begin{align}
\label{eq:barN}
\bar{N}_a:=\sqrt{x_a^2+y_a^2},
\qquad
\alpha_n^\prime \equiv \frac{d\alpha_n}{d\left(\Lambda^{-2}A^2\right)},
\qquad \alpha_n^{\prime\prime} \equiv \frac{d^2\alpha_n}{d\left(\Lambda^{-2}A^2\right)^2} \qquad \forall n=0,1,2.
\end{align}

Notice that, while the reality of $A_\mu$ implies
that of $[\chi^2]$,
this is not the case for the branch-dependent trace $[\chi]$.
Indeed, for $[\chi]\in\mathbb{R}-\{0\}$ to hold true, the inequality constraint
\begin{align}
\label{eq:ineqcons}
\mathbf{A}+\mathbf{B}+2(-1)^a \mathbb{D}= N_a^2 >0
\end{align}
must be fulfilled. Observe that the inequality cannot be saturated, owing to (\ref{eq:req}).
Important and non-trivial as the on-shell enforcement of (\ref{eq:ineqcons}) is,
we are about to show that more pressing subtleties of two-dimensional Proca-Nuevo must be considered first.
In fact, we will show that there is no need to worry over (\ref{eq:ineqcons}).

To end this section, we point out that few but consequential works
exist that adopt a homologous viewpoint to ours, albeit in the context of four-dimensional massive gravity theories whose action depends on the square root of some matrix.
Namely, works that deal with formal aspects and phenomenological implications of solving a befitting instance of our equation (\ref{eq:matrix2}). 
We find four examples particularly worthy of mention.
In chronological order, they are as follows.
First, the unavoidable (yet not necessarily physically pathological) singularities in the determinant of the root matrix that arise during dynamical evolution~\cite{Gratia:2013uza}.
Second, the (un)feasibility of the physically inequivalent branches at the perturbative level~\cite{Comelli:2015ksa}.
Third, the break-down of standard perturbation theory in certain branches
and an innovative method for its restoration, at the expense of introducing non-analyticity~\cite{Golovnev:2017lqm}.
Fourth, the physical inviability of all but one branch in bimetric gravity~\cite{Hassan:2017ugh}.
To arrive at such neat conclusion, the last reference implements the appropriate reality conditions --- counterparts to our (\ref{eq:ineqcons}).

\section{Constraint analysis}
\label{sec:cons}

Next, we make use of the Lagrangian algorithm put forward in~\cite{ErrastiDiez:2020dux} so as to elucidate the constraint structure of the two branches of two-dimensional Proca-Nuevo.
Among other results, we will thus provide an explicit count of the degrees of freedom being propagated\footnote{Sensu stricto, field theories possess infinite degrees of freedom. Here, we align ourselves with the vast literature that understands degrees of freedom in a field theory as half the number of initial conditions that must be specified in order to define the associated Cauchy problem.}.
Our starting point is the Lagrangian density (\ref{eq:2dLagfin}).

\vspace*{0.5cm}

\noindent
\textbf{Primary stage.}\\
The Euler-Lagrange equations for $(A_0,A_1)$ follow from 
the principle of stationary action.
By definition of Proca-Nuevo, these are second-order partial differential equations, linear in the generalized accelerations $(\ddot{A}_0,\ddot{A}_1)$.
Mathematically,
\begin{align}
\label{eq:ELeqs}
W^{\mu\nu}\ddot{A}_\nu+u^\mu=0,
\end{align}
where the primary Hessian $W^{\mu\nu}$ is 
\begin{align}
\label{eq:primW}
W^{\mu\nu}_{(a)}=-\frac{\alpha_1}{\Lambda^4N_a^3}\left(
\begin{array}{ccccc}
y_a^2 & (-1)^a x_a y_a \vspace*{0.3cm}\\
(-1)^a x_ay_a & x_a^2
\end{array}
\right).
\end{align}
The explicit form of the acceleration-independent piece $u^\mu$ is relegated to the appendix, (\ref{eq:primu0}) and (\ref{eq:primu1}).

Three relevant remarks are due. 
First, the linear independence between $W_{(1)}$ and $W_{(2)}$ signals the factual physical inequivalence of the two branches in the setup.
Second, for the choices (\ref{eq:exchoices}) in the second branch, the above primary Hessian reproduces that of the example in~\cite{deRham:2020yet}. 
Third, the primary Hessian is both symmetric, $W=W^T$, and generalized idempotent, $W^2\propto W$. The former feature follows by definition. The latter feature entails
the generalized idempotency of its Moore-Penrose pseudo-inverse --- shown in the appendix (\ref{eq:MRpi}) ---
and gives rise to remarkable properties for $W$~\cite{MPpi}.
The two features together imply that $W$ is a generalized projection matrix.

For both physically inequivalent branches, it is straightforward to see that the primary Hessian (\ref{eq:primW}) is singular:
$\textrm{det}(W^{\mu\nu})=0$,
which indicates that two-dimensional Proca-Nuevo is a class of constrained theories.
Further, it is evident that $W^{\mu\nu}$ has rank one,
which implies that, up to an irrelevant sign,
it admits a single normalized null-vector
\begin{align}
\label{eq:null}
\gamma_\mu^{(a)}=\frac{1}{\bar{N}_a}\Big(-(-1)^ax_a, \,\,y_a\Big).
\end{align}
We notice that, given (\ref{eq:exchoices}) for $a=2$, the above null-vector matches that of the example in~\cite{deRham:2020yet}, up to normalization.
No more comparisons between our general scenario and the example in~\cite{deRham:2020yet}
will be possible, since explicit calculations for the latter cease at precisely this point.
We will come back to this observation.

According to our above discussion,
there exists a unique primary constraint $\varphi=0$,
which defines
the primary constraint surface $T\mathcal{C}_1$ within
the tangent bundle $T\mathcal{C}$ of two-dimensional Proca-Nuevo's configuration space $\mathcal{C}=\textrm{span}\{A_0,A_1\}$.
We express this as
$\varphi\underset{1}{:\approx}0$.
Here, $\varphi$ follows from the contraction of the null-vector
(\ref{eq:null}) with the Euler-Lagrange equations (\ref{eq:ELeqs}): 
$\varphi\equiv\gamma_\mu u^\mu$.
Without loss of generality and for simplicity, we omit
an overall factor $(\Lambda^2\bar{N}_a)^{-1}$ and find
\begin{align}
\label{eq:primcons}
\varphi^{(a)}=
\displaystyle
\alpha_1\Phi
+\alpha_0^\prime \phi_0
+\alpha_1^\prime \phi_1
+\alpha_2^\prime\phi_2
\underset{1}{:\approx}0,
\end{align}
where we have introduced
\begin{align}
\label{eq:defphis}
\begin{array}{lllll}
&\displaystyle
\hspace*{-0.3cm}\Phi:=\frac{1}{N_a}\left(y_a\partial_1x_a-x_a\partial_1y_a\right), 
&\phi_0:=-2\left[(-1)^ax_aA_0+y_aA_1\right], \vspace*{0.3cm} \\
&\hspace*{-0.3cm}\phi_1:=-N_a\left\{\left[(-1)^aX+2\right]A_0-(-1)^aYA_1\right\},
&\phi_2:=-4\left[(-1)^aXx_a+Yy_a\right]A_0+4\left[(-1)^aXy_a+Yx_a\right]A_1,
\end{array}
\end{align}
which further depend on 
\begin{align}
\label{eq:defXY}
X:=x_1+x_2, \qquad Y:=y_1-y_2.
\end{align}

\vspace*{0.5cm}

\noindent
\textbf{Secondary stage.}\\
Self-consistency of the dynamics following from (\ref{eq:2dLagfin})
requires that the primary constraint
holds true under time evolution.
This means that not only $\varphi$,
but also its time derivative $\dot{\varphi}$,
{\it must} vanish in $T\mathcal{C}_1$:
\begin{align}
\label{eq:tprim}
\dot{\varphi}\underset{1}{\approx}0.
\end{align}
We stress the essentiality of (\ref{eq:tprim}),
origin of potential secondary constraints in the system.
Similarly to the Euler-Lagrange equations (\ref{eq:ELeqs}) and
in full generality, the above can be written as 
\begin{align}
\label{eq:EL2}
\widetilde{W} \gamma^\mu\ddot{Q}_\mu+\widetilde{u}\underset{1}{\approx}0,
\end{align}
where the null-vector $\gamma$ was determined in (\ref{eq:null}) and the secondary Hessian $\widetilde{W}$ is
\begin{align}
\label{eq:secHess}
\Lambda^2 \bar{N}_a \widetilde{W}_{(a)}=
\alpha_0^\prime \phi_0
+\alpha_1^\prime \phi_1
+\alpha_2^\prime\phi_2
\underset{1}{\approxident}-\alpha_1\Phi\underset{1}{\not\approxident}0.
\end{align}
The explicit form of $\widetilde{u}$ is cumbersome and we have no use for it.
Therefore, it is omitted.

In (\ref{eq:secHess}), it is of utmost importance to notice that,
for arbitrary $\alpha_n$'s and
in either of the two $T\mathcal{C}_1$-equivalent forms
in which $\widetilde{W}$ can be written, the secondary Hessian does {\it not}
identically vanish in the primary constraint surface.
This is true for both branches of two-dimensional Proca-Nuevo
--- including the explicit example in~\cite{deRham:2020yet} that
follows from setting (\ref{eq:exchoices}) in the second branch.
Being a scalar,
the secondary Hessian is thus generically non-singular.
As a result, no constraints arise at the secondary stage and 
the constraint algorithm closes\footnote{The interested reader is
referred to the discussion and example on {\it dynamical closure} in~\cite{ErrastiDiez:2020dux}.}.

\vspace*{0.5cm}

\noindent
\textbf{Degrees of freedom.}\\
For non-gauge field theories, such as those within the Proca-Nuevo class --- recall our final remark in section \ref{sec:review} ---,
the number of degrees of freedom $n_{\textrm{dof}}$ being propagated is given by~\cite{Diaz:2014yua,Diaz:2017tmy}
\begin{align}
\label{eq:dof}
n_{\textrm{dof}}=N-\frac{l}{2},
\end{align}
where $N$ is the number of a priori independent field variables
the Lagrangian depends upon and where $l$ is the number of functionally independent Lagrangian constraints (primary, plus secondary, plus tertiary, etc.).
Concerning two-dimensional Proca-Nuevo (\ref{eq:2dLagfin}), we readily see that $N=2$ (these are $A_0$ and $A_1$),
while $l=1$, which refers to (\ref{eq:primcons}).
Hence, we count $n_{\textrm{dof}}=3/2$.

In the Hamiltonian picture, the constraint
(\ref{eq:primcons}) is second class and therefore removes only half a degree of freedom.
We arrive at this conclusion by the well-established physical equivalence between
the Lagrangian and Hamiltonian formulations~\cite{Kamimura,Sugano,Pons:1986zg,Gracia}.
An explicit proof would require verifying that
the equal time Poisson bracket $\{\varphi(t,x),\varphi(t,x^\prime)\}$ does {\it not} vanish in $T\mathcal{C}_1$.
Such calculation is difficult, since the identification of Darboux coordinates for two-dimensional Proca-Nuevo
is onerous.
In fact, this hurdle partly motivates our purely Lagrangian approach
and its circumvention admits a parallel to the viewpoint advocated in~\cite{Toms:2015lza}.

The propagation of half degrees of freedom by two-dimensional Proca-Nuevo can strike one as surprising.
Indeed, in agreement with our first remark at the end of section \ref{sec:review} and in the very words of the postulators of the full class of theories, ``since we are dealing with a parity preserving Lorentz-invariance theory, there can be no half number of propagating degrees of freedom and hence the existence of a primary second class constraint automatically ensures the existence of a secondary constraint''~\cite{deRham:2020yet}.
The authors refer to~\cite{deRham:2014zqa} for further details.
Most unfortunately, we could not find an explicit proof for their claim neither in~\cite{deRham:2014zqa} nor elsewhere.
At any rate, this is the reason why the authors of~\cite{deRham:2020yet} stop their constraint analysis after inferring the existence of a primary constraint.

Certainly, the propagation of half degrees of freedom by Lorentz-violating settings is well-established.
A renowned example is Ho\v{r}ava-Lifschitz gravity~\cite{Horava:2009uw},
as explicitly shown in~\cite{Li:2009bg}
and further elucidated in~\cite{Blas:2009yd,Henneaux:2010vx}.
There exists an explicit proof~\cite{Crisostomi:2017aim}
for the avoidance of half degrees of freedom by Lorentz-invariant theories,
regardless of parity.
However, Proca-Nuevo does {\it not} fulfill the axioms of such fine proof
--- in particular, its Lagrangian does {\it not} depend on the second-order time derivatives of the vector field $\ddot{A}_\mu$. 

Likewise, parity-violating constructions are known to propagate half degrees of freedom.
Recall the archetypal example of a field theory containing a single second class constraint: a chiral or self-dual boson in two spacetime dimensions~\cite{Floreanini:1987as}.
The theory is shown to be (unorthodoxly) Lorentz-invariant from its very postulation, but a formulation that makes this feature manifest has only recently appeared~\cite{Townsend:2019koy}.
We do not have remarks beyond those in~\cite{deRham:2014zqa} regarding the relation between parity and degrees of freedom.

To the best of our knowledge, two-dimensional Proca-Nuevo is the first example of a Lorentz- and parity-invariant field theory
that is shown to propagate half degrees of freedom.

The perhaps shocking additional half degree of freedom in two-dimensional Proca-Nuevo is undeniably worrisome, because it corresponds to a so-called Boulware-Deser mode~\cite{Boulware:1972yco} --- or {\it ghost}, for short. The situation resembles the case of generic non-linear extensions of the massive gravity theory by Fierz and Pauli~\cite{Fierz:1939ix}, where constraints inherent to the linear regime no longer hold true,
leading to the propagation of degrees of freedom beyond
the usual polarizations for a massive spin two. 
It takes a non-trivial fine-tuning of such frameworks to restore all due constraints and thus avoid ghost modes.
A popular such example was put forward by de Rham, Gabadadze and Tolley (dRGT) in~\cite{deRham:2010ik,deRham:2010kj}.
Shortly after, the secondary constraints that render dRGT ghostfree were explicitly obtained~\cite{Hassan:2011vm,Hassan:2011hr}.
As a matter of fact, the ghostfreedom of dRGT has been established in several manners and also in more general frameworks, e.g~\cite{deRham:2011qq,Hassan:2011tf,Hassan:2011ea,Mirbabayi:2011aa,Golovnev:2011aa,Hassan:2012qv,Deffayet:2012nr}.

Another comparison, to the multi-field Generalized Proca (multi-GP) family of theories, is even more apposite. True to its name, multi-GP
is the exhaustive multi-field extension of Generalized Proca.
These theories generically possess a non-singular secondary Hessian, which implies the absence of secondary constraints
and thus leads to the propagation of a Boulware-Deser type of half degree of freedom per vector field under consideration.
Namely, the exact situation we have encountered for two-dimensional Proca-Nuevo.
Even so, the functional freedom in the multi-GP Lagrangian can be used to identify
a subset of ghostfree interactions: those fulfilling the {\it secondary constraint enforcing relations} unveiled in~\cite{ErrastiDiez:2019ttn,ErrastiDiez:2019trb}. Such relations uniquely ensure the appropriate singularity of the secondary Hessian,
which automatically begets a functionally independent secondary constraint per vector field.
The constraint algorithm closes in the next stage, via a non-singular tertiary Hessian and so ghosts are avoided\footnote{Remarkably, the single-field limit (i.e. Generalized Proca)
does not suffer from this problem.
In this case, the secondary Hessian is a scalar that vanishes automatically, the secondary constraint follows spontaneously and the algorithm closes with a non-singular tertiary Hessian. It is in this manner that Generalized Proca is ghostfree.}. 
For a short yet comprehensive review of multi-GP the reader may consult~\cite{ErrastiDiez:2021ykk}.

Drawing inspiration from ghostfree massive gravity theories and multi-GP,
we proceed to investigate whether the free functions 
in the Lagrangian (\ref{eq:2dLagfin}) can be restricted so as to render two-dimensional Proca-Nuevo ghostfree.

\section{No ghostfree limit}
\label{sec:limit}

In view of the general results obtained in the previous section \ref{sec:cons}, we now turn to finding those
$(\alpha_0,\alpha_1,\alpha_2)$, if any, that prevent the propagation of a ghost by two-dimensional Proca-Nuevo.

First, observe that,
in order for the primary Hessian (\ref{eq:primW})
to possess the essential rank one feature, it is necessary that 
\begin{align}
\label{eq:alpha1not0}
\alpha_1\neq 0. 
\end{align}
Then, the singularity of the secondary Hessian (\ref{eq:secHess}) when evaluated in $T\mathcal{C}_1$
can only be enforced by requiring
\begin{align}
\label{eq:noprime}
\alpha_0^\prime=0, \qquad
\alpha_1^\prime=0, \qquad
\alpha_2^\prime=0. 
\end{align}
Namely, by demanding that the $\alpha$'s in 
(\ref{eq:2dLagfin}) be constants and {\it not}
functions of the vector field squared,
in direct contradiction to (\ref{eq:alphas}), which is the original proposal in~\cite{deRham:2020yet}. 
That (\ref{eq:noprime}) is the one and only possibility follows from the fact that $(\phi_0,\phi_1,\phi_2)$ are
functionally independent from each other. To convince oneself that this is indeed the case, it suffices to check that
\begin{align}
\label{eq:detnot0}
\textrm{det}\left(
\begin{array}{cccc}
\displaystyle
\frac{\partial \phi_n}{\partial \dot{A}_0} &\qquad
\displaystyle
\frac{\partial \phi_n}{\partial \dot{A}_1} \vspace*{0.3cm}\\
\displaystyle
\frac{\partial \phi_m}{\partial \dot{A}_0} &\qquad
\displaystyle
\frac{\partial \phi_m}{\partial \dot{A}_1}
\end{array}
\right)\neq 0 \qquad \forall n,m=0,1,2 \quad \textrm{with } n\neq m.
\end{align}

When both (\ref{eq:alpha1not0}) and (\ref{eq:noprime}) are fulfilled, the primary constraint
of two-dimensional Proca-Nuevo simplifies notably. Instead of (\ref{eq:primcons}), one finds
\begin{align}
\label{eq:primok}
\varphi^{(a)}=\alpha_1\Phi\underset{1}{:\approx}0.
\end{align}
Its stability condition (\ref{eq:EL2})
also reduces dramatically, to just
the 
acceleration-independent piece $\widetilde{u}$,
which thus coincides with the secondary constraint
$\widetilde{\varphi}$
in the setup.
At this point and for both physically inequivalent branches, something peculiar happens:
the secondary constraint trivializes.
Namely, it turns out to be proportional to the primary constraint and so it identically vanishes in the first constraint surface $T\mathcal{C}_1$:
\begin{align}
\label{eq:tildeu}
\widetilde{\varphi}^{(a)}=\widetilde{u}^{(a)}=\kappa \varphi^{(a)}\underset{1}{\approxident}0,
\end{align}
where the proportionality factor $\kappa$ is explicitly shown in the appendix (\ref{eq:kappa}).
The above implies that a (functionally independent) secondary constraint {\it never} exists for two-dimensional Proca-Nuevo.
Moreover, (\ref{eq:tildeu}) indicates the closure of the constraint algorithm\footnote{For more details, the reader can consult
the discussion and example on {\it non-dynamical closure} of type A in~\cite{ErrastiDiez:2020dux}.}.

As in the general case, we again count $n_{\textrm{dof}}=3/2$, where $l=1$ now refers to the primary constraint (\ref{eq:primok}).
In conclusion, two-dimensional Proca-Nuevo remains ghostful even when the healthy rank reduction
of the primary and secondary Hessians is enforced via (\ref{eq:alpha1not0}) and (\ref{eq:noprime}).

To better understand the unexpected situation, let us elaborate on the parallel between multi-GP and Proca-Nuevo.
For $M$ number of GP fields, a certain set of conditions on the Lagrangian uniquely guarantees the necessary rank reduction of the primary Hessian
{\it and} the generation of $M$ functionally independent primary constraints.
A further set of unique conditions is needed to replicate the situation at the secondary stage: appropriate rank reduction of the secondary Hessian
{\it and} generation of $M$ functionally independent secondary constraints.
Then, the algorithm closes. In the language of~\cite{ErrastiDiez:2019ttn,ErrastiDiez:2019trb},
these conditions are the {\it primary} and {\it secondary constraint enforcing relations}, respectively. 
Moving to Proca-Nuevo, we can immediately regard (\ref{eq:alpha1not0}) as the primary constraint enforcing relation.
However, (\ref{eq:noprime}) is {\it not} a secondary constraint enforcing relation:
it does lead to the correct rank reduction of the secondary Hessian, but it does {\it not} generate a secondary constraint.
Instead, it produces the identical vanishing in $T\mathcal{C}_1$ of the would-be secondary constraint and in this way the algorithm closes.
We propose to think of (\ref{eq:noprime}) as a {\it secondary constraint trivializing relation}.

\section{Incurability}
\label{sec:rigid}

Two-dimensional Proca-Nuevo is incurable, in the sense that its extension to incorporate interactions belonging to other electrodynamics theories in table \ref{table:before} remains a ghostful setup. Elucidating and proving this claim is the aim of the present section.

\vspace*{0.5cm}

\noindent
\textbf{Incurability via massless electrodynamics.}\\
Consider supplementing (\ref{eq:2dLagfin}) with Maxwell electrodynamics
\begin{align}
\label{eq:Max}
S_{\textrm{M}}=\int_{\mathcal{M}}d^2x\,\, \mathcal{L}_{\textrm{M}}, \qquad
\mathcal{L}_{\textrm{M}}=-\frac{1}{4}F_{\mu\nu}F^{\mu\nu}=\frac{\Lambda^4}{2}y_2^2.
\end{align}
Then, the primary Hessian (\ref{eq:primW})
receives the additional contribution
\begin{align}
\label{eq:HessMax}
\left(
\begin{array}{ccc}
0 &\quad 0 \\
0 &\quad 1
\end{array}
\right),
\end{align}
which results in a non-singular total primary Hessian.
In this case, the system has no constraints at all ($l=0$) and thus propagates two degrees of freedom $n_{\textrm{dof}}=2$, one of which is a ghost.

By definition, the same holds true for any electrodynamics theory within the NLE class, whose primary Hessian is of the form (\ref{eq:HessMax}),
albeit with a more complicated, theory-dependent non-zero entry.

\newpage

\noindent
\textbf{Incurability via massive electrodynamics.}\\
Consider now two-dimensional Proca-Nuevo supplemented by the standard hard mass term
of Proca electrodynamics, 
\begin{align}
\label{eq:Pr}
S_{\textrm{P}}=\int_{\mathcal{M}}d^2x\,\, \mathcal{L}_{\textrm{P}}, \qquad
\mathcal{L}_{\textrm{P}}=-\frac{1}{2}m^2A_\mu A^\mu, \qquad m\in\mathbb{R}^+.
\end{align}
It is a direct consequence of our analyses in sections \ref{sec:cons} and \ref{sec:limit} that the resulting setup is ghostful.

Next, we discuss the incompatibility with the Self-interacting Proca subset of Generalized Proca:
$\mathcal{L}_2$ in equation (12) of~\cite{BeltranJimenez:2016rff}.
Those terms within Self-interacting Proca that do not depend on the field strength $F_{\mu\nu}$
follow the same logic as Proca electrodynamics: they
are expressly shown to yield a ghost by our proofs in sections \ref{sec:cons} and \ref{sec:limit}.
For the complementary subset --- Self-interacting Proca depending on $F_{\mu\nu}$ ---, Lorentz-invariance dictates the appearance of at least two field strengths.
This structure immediately lands us back into the previous, massless-like scenario: disruption of the singularity of the primary Hessian
and propagation of two degrees of freedom, one of which is a ghost.

By definition, Generalized Proca is a class of Lorentz-invariant, first-order electrodynamics theories
that possesses the same constraint structure as Proca electrodynamics~\cite{ErrastiDiez:2019ttn,ErrastiDiez:2019trb}.
Consequently, the addition of self-interaction terms of the Generalized Proca class (beyond Self-interacting Proca) to Proca-Nuevo
will generically prevent the generation of a secondary constraint,
through the non-degeneracy of the secondary Hessian.
Namely, a more involved instance of (\ref{eq:secHess}) will take place.
This entails the generic propagation of a ghost mode.

Extended Proca-Nuevo~\cite{deRham:2021efp} is precisely the result of supplementing Proca-Nuevo with certain terms of the Generalized Proca class. 
By construction, this proposal is generically ghostful.
In search of completeness, two-dimensional
Extended Proca-Nuevo is explicitly shown to be {\it irremediably} ghostful in the appendix, (\ref{eq:LagEPN2d})-(\ref{eq:betap0}). 

\section{Conclusions}
\label{sec:concl}

For two-dimensional Proca-Nuevo in the convenient form (\ref{eq:2dLagfin}), we have
explicitly obtained the only primary constraint (\ref{eq:primcons})
and demonstrated that, in general, it does {\it not} lead to a secondary constraint (\ref{eq:secHess}).
The implication is that two-dimensional Proca-Nuevo is generically ghostful. 
Remarkably, the necessary and sufficient conditions (\ref{eq:alpha1not0}) and (\ref{eq:noprime})
that lay the ground for the generation of a secondary constraint fail to accomplish their life purpose.
This is because the would-be secondary constraint trivializes in the first constraint surface (\ref{eq:tildeu}). 
As a result, there is no restriction possible on the free functions of two-dimensional Proca-Nuevo that renders the setup ghostfree. 
Two-dimensional Proca-Nuevo is thus irremediably ghostful
and, as far as we know, the first example
of a Lorentz- and parity-invariant theory propagating half degrees of freedom.
Last but not least, we have shown that the construction cannot be cured
via the addition of terms from other electrodynamics theories in table \ref{table:before}. 
Such generalizations, which encompass Extended Proca-Nuevo, are unavoidably ghostful themselves.

Summing up, for two spacetime dimensions, we have shown that table \ref{table:before} and {\it not} table \ref{table:after}
is the correct classification of electrodynamics theories within scope. 

\vspace*{0.5cm}

\noindent
\textbf{Lessons for higher dimensions.}\\
Low dimensional toy-models of physically relevant four-dimensional field theories
can provide useful information and be amenable to otherwise unfeasible or even impossible calculations.
The case of color confinement in Quantum ChromoDynamics (QCD) is paradigmatic, e.g.~\cite{tHooft:1974pnl,Witten:1978ka}.
However, these auxiliary frameworks call for extreme caution in their interpretation, so as to avoid wrong extrapolations.
Two-dimensional General Relativity
is an eloquent example of a toy-model with
a crooked relationship to its higher-dimensional counterparts,
as the lucid studies~\cite{Lemos:1993hz,Deser:1995cs} demonstrate.

It is a fact that the constraint structures of all electrodynamics theories in table \ref{table:before}
are insensitive to the (finite) dimension $d\geq 2$ of the underlying Minkowski spacetime $\mathcal{M}$.
In other words, 
such massless (massive) electrodynamics theories invariably have two first (second) class constraints and so they propagate $n_{\textrm{dof}}=d-2$ ($n_{\textrm{dof}}=d-1$) degrees of freedom on $\mathcal{M}$.
We stress that nothing dramatic happens to massless electrodynamics for $d=2$: they simply trivialize\footnote{For elaborate and clever presentations of two-dimensional Maxwell electrodynamics, the reader can consult the lecture notes by Wheeler~\cite{Wheeler}
or McDonald~\cite{McDonald}.}.
Besides, there exists a notable parallel in the construction of Generalized Proca and Proca-Nuevo
in terms of elementary symmetric polynomials --- compare (\ref{eq:mathlps})-(\ref{eq:elempols}) and (\ref{eq:ESpSEPN}).
Both considerations are by no means enough to conclude that (Extended) Proca-Nuevo is generically, let alone irremediably, ghostful in $d>2$.

Nonetheless, valuable insights can be extracted from our work.
Two-dimensional Proca-Nuevo
constitutes a counterexample to the lore that
considers simultaneous Lorentz and parity invariances
as sufficient conditions for the generation of secondary constraints
in a system with primary constraints.
In full generality, the inference of a primary second class constraint falls short from ensuring the existence of a secondary second class constraint.
Our analyzed toy-models thus urge to exercise prudence
as regards the ghostfreedom claims for (Extended) Proca-Nuevo in four dimensions. 
The impossibility in section \ref{sec:limit} to find even a restricted functional space
that will accomplish the sought secondary constraint
elongates the shadow of doubt.
Both classes of non-linear massive electrodynamics require further investigation in this respect.
As of currently, we sustain that there is no unequivocal
identification for the correct classification of relevant electrodynamics among tables \ref{table:before} and \ref{table:after}.


\vspace*{0.5cm}

\noindent
{\bf Acknowledgments.}
We are deeply indebted to Julio A. M\'endez-Zavaleta for his painstaking and constructive skepticism
over a series of insightful discussions at the final stage of this work, and for bringing up~\cite{Hassan:2017ugh}.
We have greatly benefited from sagacious correspondence and conversations with Mar\'ia Jos\'e Guzm\'an.
We thank Nicol\'as Coca-L\'opez and Markus Maier for their careful readings of earlier versions
of this manuscript, which have led to a significant improvement in the writing.
This work is funded by the Deutsche Forschungsgemeinschaft (DFG, German Research Foundation) under Germany's Excellence Strategy -- EXC-2094 -- 390783311.

\newpage 

\appendix

\section{Supplementary formulae}
\label{app:math}

In this appendix, we provide auxiliary results,
for the benefit of the reader interested in reproducing our calculations. 

\vspace*{0.5cm}

\noindent
\textbf{Two-dimensional Proca-Nuevo.}\\
We begin with the acceleration independent piece $u^\mu$
of the Euler-Lagrange equations (\ref{eq:ELeqs})
for the double-branched, two-dimensional Proca-Nuevo theory (\ref{eq:2dLagfin}).
The time component is
\begin{align}
\label{eq:primu0}
\begin{array}{lllll}
\Lambda^4 u^0_{(a)}&\hspace*{-0.2cm}=&\hspace*{-0.2cm}
\displaystyle
\frac{\alpha_1}{N_a^3}\left[x_ay_a\left(2\partial_1\dot{A}_0-(-1)^a\partial_1^2A_1\right)-x_a^2\partial_1^2A_0+(-1)^a\bar{N}_a^2\partial_1\dot{A}_1\right]  \vspace*{0.3cm} \\
&\hspace*{-0.2cm} &\hspace*{-0.2cm} \displaystyle
+2\left[(-1)^a\frac{\alpha_1^\prime}{N_a} y_a+2\alpha_2^\prime\left(y_1+y_2\right)\right]
\left(-A_0\partial_1 A_0+A_1\partial_1A_1\right) \vspace*{0.3cm} \\
&\hspace*{-0.2cm} &\hspace*{-0.2cm} \displaystyle
-2\left[(-1)^a\frac{\alpha_1^\prime}{N_a} x_a+2\alpha_2^\prime\left(x_1+x_2\right)\right]
\left(-A_0\dot{A}_0+A_1\dot{A}_1\right) \vspace*{0.3cm} \\
&\hspace*{-0.2cm} &\hspace*{-0.2cm} \displaystyle
+2\Lambda^2\left[\alpha_0^\prime+\alpha_1^\prime N_a+\alpha_2^\prime\left(N_2^2-N_1^2\right)\right]A_0,
\end{array}
\end{align}
while the space component is
\begin{align}
\label{eq:primu1}
\begin{array}{lllll}
\Lambda^4 u^1_{(a)}&\hspace*{-0.2cm}=&\hspace*{-0.2cm}
\displaystyle
\frac{\alpha_1}{N_a^3}\left[x_ay_a\left(2\partial_1\dot{A}_1-(-1)^a\partial_1^2A_0\right)-y_a^2\partial_1^2A_1+(-1)^a\bar{N}_a^2\partial_1\dot{A}_0\right]  \vspace*{0.3cm} \\
&\hspace*{-0.2cm} &\hspace*{-0.2cm} \displaystyle
+2\left[\frac{\alpha_1^\prime}{N_a} x_a-2\alpha_2^\prime\left(x_1-x_2\right)\right]
\left(-A_0\partial_1 A_0+A_1\partial_1A_1\right) \vspace*{0.3cm} \\
&\hspace*{-0.2cm} &\hspace*{-0.2cm} \displaystyle
-2\left[\frac{\alpha_1^\prime}{N_a} y_a-2\alpha_2^\prime\left(y_1-y_2\right)\right]
\left(-A_0\dot{A}_0+A_1\dot{A}_1\right) \vspace*{0.3cm} \\
&\hspace*{-0.2cm} &\hspace*{-0.2cm} \displaystyle
-2\Lambda^2\left[\alpha_0^\prime+\alpha_1^\prime N_a+\alpha_2^\prime\left(N_2^2-N_1^2\right)\right]A_1.
\end{array}
\end{align}

Next, we show the Moore-Penrose pseudo-inverse $M$ of the primary Hessian $W$ in (\ref{eq:primW}). As it turns out, $M$ is proportional to $W$:
\begin{align}
\label{eq:MRpi}
M_{(a)}= \left(\frac{\Lambda^4 N_a^3}{\alpha_1\bar{N}_a^2} \right)^2 W_{(a)},
\end{align}
which follows from the symmetry and generalized idempotency of $W$ noted below (\ref{eq:primW}).

On the other hand, the (vector field dependent) proportionality factor $\kappa$ in (\ref{eq:tildeu}) is
\begin{align}
\label{eq:kappa}
\kappa=\frac{2x_ay_a}{\Lambda^2N_a^2\bar{N}_a^2}\left[-x_a\partial_1 x_a+y_a\partial_1y_a+N_a^2\partial_1\delta(x-x^\prime)\right].
\end{align}
$\kappa$ linearly relates the (would-be) secondary constraint to the primary constraint when Proca-Nuevo is subjected to the conditions (\ref{eq:alpha1not0})
and (\ref{eq:noprime}).

\vspace*{0.5cm}

\noindent
\textbf{Two-dimensional Extended Proca-Nuevo.}\\
Turning to Extended Proca-Nuevo,
the prescription in~\cite{deRham:2021efp}
is to add together Proca-Nuevo and those Generalized Proca interactions that leave the primary Hessian of Proca-Nuevo unchanged, while allowing for distinct vector field squared dependent coefficients to precede each functionally independent term.
In two dimensions, this means enlarging
(\ref{eq:2dLagfin}) to 
\begin{align}
\label{eq:LagEPN2d}
\mathcal{L}^{(a)}_{\textrm{EPN}_2}=
\alpha_0+\alpha_1N_a+\alpha_2\sum_{a=1}^2(-1)^aN_a^2+\sum_{n=0}^{2}\beta_n\mathcal{L}_n(\partial_\mu A_\nu),
\end{align}
where $(\alpha,\beta)$'s are arbitrary smooth functions of the (suitably normalized) vector field squared: (\ref{eq:alphas}) and
\begin {align}
\label{eq:betas}
\beta_n=\beta_n\left(\frac{A_\mu A^\mu}{\Lambda^2}\right)\equiv\beta_n\left(\frac{A^2}{\Lambda^2}\right)
\qquad \forall n=0,1,2,
\end{align}
and where $\mathcal{L}_n(\partial_\mu A_\nu)$ are the univariate ESPs of degree $n$, explicitly given by
\begin{align}
\label{eq:ESpSEPN}
\begin{array}{llll}
&\displaystyle \mathcal{L}_0(\partial_\mu A_\nu)=1, \qquad \qquad
\mathcal{L}_1(\partial_\mu A_\nu)=\frac{1}{\Lambda^2}\partial_\mu A^\mu=x_2-2, \vspace*{0.3cm}\\
&\displaystyle \mathcal{L}_2(\partial_\mu A_\nu)=\frac{1}{\Lambda^4}\left(\partial_\mu A^\mu\partial_\nu A^\nu-\partial_\mu A_\nu \partial^\nu A^\mu\right)=-N_1^2+N_2^2-2(x_2-2).
\end{array}
\end{align}
For simplicity and without loss of generality,
we redefine
\begin{align}
\label{eq:redef}
\alpha_0^{(\textrm{new})}= \alpha_0+\beta_0-2\beta_1+4\beta_2, \qquad 
\alpha_2^{(\textrm{new})}= \alpha_2+\beta_2, \qquad
\beta^{(\textrm{new})}= \beta_1-2\beta_2,
\end{align}
drop the ``new'' labels and obtain
\begin{align}
\label{eq:LagEPN2dfin}
\mathcal{L}^{(a)}_{\textrm{EPN}_2}=
\alpha_0+\alpha_1N_a+\alpha_2\sum_{a=1}^2(-1)^aN_a^2+\beta x_2.
\end{align}
Notice that the non-triviality of the new term, as compared with (\ref{eq:2dLagfin}), 
necessitates that $\beta$ depends on the vector field.

Next, we repeat the constraint analysis in sections \ref{sec:cons} and \ref{sec:limit}
for the full Extended Proca-Nuevo (\ref{eq:LagEPN2dfin}).

\vspace*{0.5cm}

\noindent
\textit{Primary stage.}
By definition, the primary Hessian (\ref{eq:primW}) is not affected by the new derivative self-interaction term.
Hence, its normalized null-vector (\ref{eq:null})
also remains unaltered. 
Contrastingly, the acceleration independent part
of the Euler-Lagrange equations
$u^\mu$ in (\ref{eq:ELeqs}) does change.
Specifically, (\ref{eq:primu0}) and (\ref{eq:primu1}) acquire the new contributions
\begin{align}
\label{eq:u0extra}
u^0=\frac{\beta^\prime}{\Lambda^2}
\left[(X-2)A_0-\bar{Y}A_1\right], \qquad
u^1=\frac{\beta^\prime}{\Lambda^2}
\left[-YA_0+\bar{X}A_1\right],
\end{align}
respectively,
where we have introduced
\begin{align}
\label{eq:defXYbar}
\beta^\prime\equiv\frac{d\beta}{d(\Lambda^{-2}A^2)}, \qquad
\bar{X}:=x_1-x_2-2, \qquad 
\bar{Y}:=y_1+y_2.
\end{align}
As a result, the  
primary constraint (\ref{eq:primcons})
also receives an additional contribution,
explicitly given by
\begin{align}
\label{eq:varphiextra}
\varphi^{(a)}=\beta^\prime\psi, \qquad
\psi:= -\left[(-1)^a(X+2)x_a+Yy_a\right]A_0+
\left[(-1)^a\bar{Y}x_a+\bar{X}y_a\right]A_1.
\end{align}

\vspace*{0.5cm}

\noindent
\textit{Secondary stage.} The necessary stability of the enlarged primary constraint
under time evolution (\ref{eq:tprim})
gives rise to the secondary Euler-Lagrange equations (\ref{eq:EL2}).
The secondary Hessian in this case is amplified by the term
\begin{align}
\label{eq:Wprimextra}
\Lambda^2\bar{N}_a\widetilde{W}=
\beta^\prime\psi.
\end{align}
Taking into consideration the functional independence among all $\alpha$'s (\ref{eq:detnot0})
and upon verifying the functional independence of $\psi$
as well,
\begin{align}
\label{eq:detnot0psi}
\textrm{det}\left(
\begin{array}{cccc}
\displaystyle
\frac{\partial \psi}{\partial \dot{A}_0} &\qquad
\displaystyle
\frac{\partial \psi}{\partial \dot{A}_1} \vspace*{0.3cm}\\
\displaystyle
\frac{\partial \phi_n}{\partial \dot{A}_0} &\qquad
\displaystyle
\frac{\partial \phi_n}{\partial \dot{A}_1}
\end{array}
\right)\neq 0 \qquad \forall n=0,1,2,
\end{align}
we conclude that both (\ref{eq:noprime}) and
\begin{align}
\label{eq:betap0}
\beta^\prime=0
\end{align}
are needed in order to try to generate a secondary constraint
that avoids the propagation of ghosts in two-dimensional Extended Proca-Nuevo.
However, (\ref{eq:betap0}) renders the extra interaction in Extended Proca-Nuevo (\ref{eq:LagEPN2dfin}), as compared to Proca-Nuevo (\ref{eq:2dLagfin}), trivial: it can be written as a boundary term.
Hence, there is no ghostfree Extended Proca-Nuevo in two-dimensions.

As a final remark, we point out that the authors of~\cite{deRham:2021efp}, who work in four spacetime dimensions,
explicitly indicate the necessity of vector field dependent $\beta$'s for ``non-trivial phenomenological effects'' in Extended Proca-Nuevo. 




\begin{thebibliography}{99}

\bibitem{Maxwell}
J.~C.~Maxwell,
``A dynamical theory of the electromagnetic field,''
Philosophical Transactions of the Royal Society {\bf 155}, 459-512 (1865)
doi:10.1098/rstl.1865.0008.

\bibitem{Pod1}
B.~Podolsky,
``A Generalized Electrodynamics Part I -- Non-Quantum,''
Phys. Rev. {\bf 62}, 68 (1942)
doi:10.1103/PhysRev.62.68.

\bibitem{Pod2}
B.~Podolsky and C.~Kikuchi,
``A Generalized Electrodynamics Part II -- Quantum,''
Phys. Rev. {\bf 65}, 228 (1944)
doi:10.1103/PhysRev.65.228.

\bibitem{Yang:1954ek}
C.~N.~Yang and R.~L.~Mills,
``Conservation of Isotopic Spin and Isotopic Gauge Invariance,''
Phys. Rev. \textbf{96}, 191-195 (1954)
doi:10.1103/PhysRev.96.191

\bibitem{Schwarz:1978cn}
A.~S.~Schwarz,
``The Partition Function of Degenerate Quadratic Functional and Ray-Singer Invariants,''
Lett. Math. Phys. \textbf{2}, 247-252 (1978)
doi:10.1007/BF00406412.

\bibitem{Kostelecky:2002hh}
V.~A.~Kostelecky and M.~Mewes,
``Signals for Lorentz violation in electrodynamics,''
Phys. Rev. D \textbf{66}, 056005 (2002)
doi:10.1103/PhysRevD.66.056005
[arXiv:hep-ph/0205211 [hep-ph]].

\bibitem{Mashhoon:2004rz}
B.~Mashhoon,
``Nonlocal electrodynamics of linearly accelerated systems,''
Phys. Rev. A \textbf{70}, 062103 (2004)
doi:10.1103/PhysRevA.70.062103
[arXiv:hep-th/0407278 [hep-th]].

\bibitem{Mashhoon:2005ft}
B.~Mashhoon,
``Nonlocal electrodynamics of rotating systems,''
Phys. Rev. A \textbf{72}, 052105 (2005)
doi:10.1103/PhysRevA.72.052105
[arXiv:hep-th/0503205 [hep-th]].

\bibitem{Mashhoon:2007sf}
B.~Mashhoon,
``Nonlocal electrodynamics of accelerated systems,''
Phys. Lett. A \textbf{366}, 545-549 (2007)
doi:10.1016/j.physleta.2007.02.071
[arXiv:hep-th/0702074 [hep-th]].

\bibitem{Chern:1974ft}
S.~S.~Chern and J.~Simons,
``Characteristic forms and geometric invariants,''
Annals Math. \textbf{99}, 48-69 (1974)
doi:10.2307/1971013.

\bibitem{Fonseca:2010av}
M.~V.~S.~Fonseca and A.~A.~Vargas-Paredes,
``Is it possible to accommodate massive photons in the framework of a gauge-invariant electrodynamics?,''
Braz. J. Phys. \textbf{40}, 319 (2010)
doi:10.1590/S0103-97332010000300011
[arXiv:1005.3480 [hep-th]].

\bibitem{Bovy}
J.~Bovy,
``The self-energy of the electron: a quintessential problem in the development of QED,''
[arXiv:physics/0608108 [physics.hist-ph].

\bibitem{Johnson}
K.~Johnson, M.~Baker and R.~Willey,
``Self-Energy of the Electron,''
Phys. Rev. {\bf 136}, B1111 (1964)
doi:10.1103/PhysRev.136.B1111.

\bibitem{Proca1}
A.~Proca,
``Sur la th\'eorie ondulatoire des \'electrons positifs et n\'egatifs,''
J. Phys. Radium {\bf 7}, 347 (1936)
doi:10.1051/jphysrad:0193600708034700.

\bibitem{Proca2}
A.~Proca,
``Th\'eorie non relativiste des particules \`a spin entier,''
J. Phys. Radium {\bf 9}, 61 (1938)
doi:10.1051/jphysrad:019380090206100.

\bibitem{Bloem}
N.~Bloembergen,
``Nonlinear Optics,''
W. A. Benjamin Inc., New York (1965).

\bibitem{Photonics}
J.~D.~Joannopoulos, S.~G.~Johnson, J.~N.~Winn and R.~D.~Meade,
``Photonic Crystals. Molding the Flow of Light,''
2nd Edition, Princeton University Press, Princeton (2008).

\bibitem{Pleb}
J.~Plebanski,
``Lectures on non-linear electrodynamics,''
NORDITA, Copenhagen (1968).

\bibitem{Born:1934gh}
M.~Born and L.~Infeld,
``Foundations of the new field theory,''
Proc. Roy. Soc. Lond. A \textbf{144}, no.852, 425-451 (1934)
doi:10.1098/rspa.1934.0059.

\bibitem{Bandos:2020jsw}
I.~Bandos, K.~Lechner, D.~Sorokin and P.~K.~Townsend,
``A non-linear duality-invariant conformal extension of Maxwell's equations,''
Phys. Rev. D \textbf{102}, 121703 (2020)
doi:10.1103/PhysRevD.102.121703
[arXiv:2007.09092 [hep-th]].

\bibitem{Kosyakov:2020wxv}
B.~P.~Kosyakov,
``Nonlinear electrodynamics with the maximum allowable symmetries,''
Phys. Lett. B \textbf{810}, 135840 (2020)
doi:10.1016/j.physletb.2020.135840
[arXiv:2007.13878 [hep-th]].

\bibitem{Sorokin:2021tge}
D.~P.~Sorokin,
``Introductory Notes on Non-linear Electrodynamics and its Applications,''
Fortsch. Phys. \textbf{70}, no.7-8, 2200092 (2022)
doi:10.1002/prop.202200092
[arXiv:2112.12118 [hep-th]].

\bibitem{Eboli:2016kko}
O.~J.~P.~\'Eboli and M.~C.~Gonz\'alez-Garc\'ia,
``Classifying the bosonic quartic couplings,''
Phys. Rev. D \textbf{93}, no.9, 093013 (2016)
doi:10.1103/PhysRevD.93.093013
[arXiv:1604.03555 [hep-ph]].

\bibitem{Freitas:2021cfi}
F.~F.~Freitas, C.~A.~R.~Herdeiro, A.~P.~Morais, A.~Onofre, R.~Pasechnik, E.~Radu, N.~Sanchis-Gual and R.~Santos,
``Ultralight bosons for strong gravity applications from simple Standard Model extensions,''
JCAP \textbf{12}, no.12, 047 (2021)
doi:10.1088/1475-7516/2021/12/047
[arXiv:2107.09493 [hep-ph]].

\bibitem{Aoki:2022woy}
K.~Aoki and M.~Minamitsuji,
``Resolving the pathologies of self-interacting Proca fields: A case study of Proca stars,''
Phys. Rev. D \textbf{106}, no.8, 084022 (2022)
doi:10.1103/PhysRevD.106.084022
[arXiv:2206.14320 [gr-qc]].

\bibitem{Tasinato:2014eka}
G.~Tasinato,
``Cosmic Acceleration from Abelian Symmetry Breaking,''
JHEP \textbf{04}, 067 (2014)
doi:10.1007/JHEP04(2014)067
[arXiv:1402.6450 [hep-th]].

\bibitem{Heisenberg:2014rta}
L.~Heisenberg,
``Generalization of the Proca Action,''
JCAP \textbf{05}, 015 (2014)
doi:10.1088/1475-7516/2014/05/015
[arXiv:1402.7026 [hep-th]].

\bibitem{Allys:2015sht}
E.~Allys, P.~Peter and Y.~Rodr\'iguez,
``Generalized Proca action for an Abelian vector field,''
JCAP \textbf{02}, 004 (2016)
doi:10.1088/1475-7516/2016/02/004
[arXiv:1511.03101 [hep-th]].

\bibitem{BeltranJimenez:2016rff}
J.~Beltr\'an Jim\'enez and L.~Heisenberg,
``Derivative self-interactions for a massive vector field,''
Phys. Lett. B \textbf{757}, 405-411 (2016)
doi:10.1016/j.physletb.2016.04.017
[arXiv:1602.03410 [hep-th]].

\bibitem{deRham:2020yet}
C.~de Rham and V.~Pozsgay,
``New class of Proca interactions,''
Phys. Rev. D \textbf{102}, no.8, 083508 (2020)
doi:10.1103/PhysRevD.102.083508
[arXiv:2003.13773 [hep-th]].

\bibitem{deRham:2021efp}
C.~de Rham, S.~Garc\'ia-S\'aenz, L.~Heisenberg and V.~Pozsgay,
``Cosmology of Extended Proca-Nuevo,''
JCAP \textbf{03}, 053 (2022)
doi:10.1088/1475-7516/2022/03/053
[arXiv:2110.14327 [hep-th]].

\bibitem{deRham:2021yhr}
C.~de Rham, L.~Heisenberg, A.~Kumar and J.~Zosso,
``Quantum stability of a new Proca theory,''
Phys. Rev. D \textbf{105}, no.2, 2 (2022)
doi:10.1103/PhysRevD.105.024033
[arXiv:2108.12892 [hep-th]].

\bibitem{Pozsgay:2022xpc}
V.~Pozsgay,
``Cosmology of a new class of massive vector fields,''
[arXiv:2203.14608 [gr-qc]].

\bibitem{deRham:2022sdl}
C.~de Rham, L.~Engelbrecht, L.~Heisenberg and A.~L\"uscher,
``Positivity bounds in vector theories,''
[arXiv:2208.12631 [hep-th]].

\bibitem{ESPs}
P.~Borwein and T.~Erd\'elyi,
``Polynomial Inequalities,''
Springer-Verlag, New York (1995).

\bibitem{Hassan:2011vm}
S.~F.~Hassan and R.~A.~Rosen,
``On Non-Linear Actions for Massive Gravity,''
JHEP \textbf{07}, 009 (2011)
doi:10.1007/JHEP07(2011)009
[arXiv:1103.6055 [hep-th]].

\bibitem{Gratia:2013uza}
P.~Gratia, W.~Hu and M.~Wyman,
``Self-accelerating Massive Gravity: Bimetric Determinant Singularities,''
Phys. Rev. D \textbf{89}, no.2, 027502 (2014)
doi:10.1103/PhysRevD.89.027502
[arXiv:1309.5947 [hep-th]].

\bibitem{Comelli:2015ksa}
D.~Comelli, M.~Crisostomi, K.~Koyama, L.~Pilo and G.~Tasinato,
``New Branches of Massive Gravity,''
Phys. Rev. D \textbf{91}, no.12, 121502 (2015)
doi:10.1103/PhysRevD.91.121502
[arXiv:1505.00632 [hep-th]].

\bibitem{Golovnev:2017lqm}
A.~Golovnev and F.~Smirnov,
``Unusual square roots in the ghostfree theory of massive gravity,''
JHEP \textbf{06}, 130 (2017)
doi:10.1007/JHEP06(2017)130
[arXiv:1704.08874 [gr-qc]].

\bibitem{Hassan:2017ugh}
S.~F.~Hassan and M.~Kocic,
``On the local structure of spacetime in ghost-free bimetric theory and massive gravity,''
JHEP \textbf{05}, 099 (2018)
doi:10.1007/JHEP05(2018)099
[arXiv:1706.07806 [hep-th]].

\bibitem{ErrastiDiez:2020dux}
V.~Errasti D\'iez, M.~Maier, J.~A.~M\'endez-Zavaleta and M.~Taslimi Tehrani,
``Lagrangian constraint analysis of first-order classical field theories with an application to gravity,''
Phys. Rev. D \textbf{102}, 065015 (2020)
doi:10.1103/PhysRevD.102.065015
[arXiv:2007.11020 [hep-th]].

\bibitem{MPpi}
O.~M.~Baksalary and G.~Trenkler,
``On matrices whose Moore-Penrose inverse is idempotent,''
Linear and Multilinear Algebra, 70:11, 2014-2026 (2020)
doi: 10.1080/03081087.2020.1781038.

\bibitem{Diaz:2014yua}
B.~D\'\i{}az, D.~Higuita and M.~Montesinos,
``Lagrangian approach to the physical degree of freedom count,''
J. Math. Phys. \textbf{55}, 122901 (2014)
doi:10.1063/1.4903183
[arXiv:1406.1156 [hep-th]].

\bibitem{Diaz:2017tmy}
B.~D\'iaz and M.~Montesinos,
``Geometric Lagrangian approach to the physical degree of freedom count in field theory,''
J. Math. Phys. \textbf{59}, no.5, 052901 (2018)
doi:10.1063/1.5008740
[arXiv:1710.01371 [gr-qc]].

\bibitem{Kamimura}
K.~Kamimura, 
``Singular Lagrangian and constrained Hamiltonian systems, generalized canonical formalism,''
Nuovo Cim. {\bf 68}, 33-54 (1982),
doi:10.1007/BF02888859.

\bibitem{Sugano}
R.~Sugano and H.~Kamo,
``Poincar\'e-Cartan Invariant Form and Dynamical Systems with Constraints,''
Progress of Theoretical Physics {\bf 68}, 1377 (1982),
doi:10.1143/PTP.67.1966.

\bibitem{Pons:1986zg}
J.~M.~Pons,
``New Relations Between Hamiltonian and Lagrangian Constraints,''
J. Phys. A \textbf{21}, 2705 (1988)
doi:10.1088/0305-4470/21/12/014.

\bibitem{Gracia}
X.~Gr\`{a}cia and J.~M.~Pons,
``Gauge generators, Dirac's conjecture, and degrees of freedom for constrained systems,''
Ann. Phys. 187, 355 (1988)
doi:10.1016/0003-4916(88)90153-4.

\bibitem{Toms:2015lza}
D.~J.~Toms,
``Faddeev-Jackiw quantization and the path integral,''
Phys. Rev. D \textbf{92}, no.10, 105026 (2015)
doi:10.1103/PhysRevD.92.105026
[arXiv:1508.07432 [hep-th]].

\bibitem{deRham:2014zqa}
C.~de Rham,
``Massive Gravity,''
Living Rev. Rel. \textbf{17}, 7 (2014)
doi:10.12942/lrr-2014-7
[arXiv:1401.4173 [hep-th]].

\bibitem{Horava:2009uw}
P.~Ho\v{r}ava,
``Quantum Gravity at a Lifshitz Point,''
Phys. Rev. D \textbf{79}, 084008 (2009)
doi:10.1103/PhysRevD.79.084008
[arXiv:0901.3775 [hep-th]].

\bibitem{Li:2009bg}
M.~Li and Y.~Pang,
``A Trouble with Ho\v{r}ava-Lifshitz Gravity,''
JHEP \textbf{08}, 015 (2009)
doi:10.1088/1126-6708/2009/08/015
[arXiv:0905.2751 [hep-th]].

\bibitem{Blas:2009yd}
D.~Blas, O.~Pujolas and S.~Sibiryakov,
``On the Extra Mode and Inconsistency of Horava Gravity,''
JHEP \textbf{10}, 029 (2009)
doi:10.1088/1126-6708/2009/10/029
[arXiv:0906.3046 [hep-th]].

\bibitem{Henneaux:2010vx}
M.~Henneaux, A.~Kleinschmidt and G.~Lucena G\'omez,
``Remarks on Gauge Invariance and First-Class Constraints,''
Proc. Steklov Inst. Math. \textbf{272}, no.1, 141 (2011)
doi:10.1134/S0081543811010123
[arXiv:1004.3769 [hep-th]].

\bibitem{Crisostomi:2017aim}
M.~Crisostomi, R.~Klein and D.~Roest,
``Higher Derivative Field Theories: Degeneracy Conditions and Classes,''
JHEP \textbf{06}, 124 (2017)
doi:10.1007/JHEP06(2017)124
[arXiv:1703.01623 [hep-th]].

\bibitem{Floreanini:1987as}
R.~Floreanini and R.~Jackiw,
``Selfdual Fields as Charge Density Solitons,''
Phys. Rev. Lett. \textbf{59}, 1873 (1987)
doi:10.1103/PhysRevLett.59.1873.

\bibitem{Townsend:2019koy}
P.~K.~Townsend,
``Manifestly Lorentz invariant chiral boson action,''
Phys. Rev. Lett. \textbf{124}, no.10, 101604 (2020)
doi:10.1103/PhysRevLett.124.101604
[arXiv:1912.04773 [hep-th]].

\bibitem{Boulware:1972yco}
D.~G.~Boulware and S.~Deser,
``Can gravitation have a finite range?,''
Phys. Rev. D \textbf{6}, 3368-3382 (1972)
doi:10.1103/PhysRevD.6.3368.

\bibitem{Fierz:1939ix}
M.~Fierz and W.~Pauli,
``On relativistic wave equations for particles of arbitrary spin in an electromagnetic field,''
Proc. Roy. Soc. Lond. A \textbf{173}, 211-232 (1939)
doi:10.1098/rspa.1939.0140.

\bibitem{deRham:2010ik}
C.~de Rham and G.~Gabadadze,
``Generalization of the Fierz-Pauli Action,''
Phys. Rev. D \textbf{82}, 044020 (2010)
doi:10.1103/PhysRevD.82.044020
[arXiv:1007.0443 [hep-th]].

\bibitem{deRham:2010kj}
C.~de Rham, G.~Gabadadze and A.~J.~Tolley,
``Resummation of Massive Gravity,''
Phys. Rev. Lett. \textbf{106}, 231101 (2011)
doi:10.1103/PhysRevLett.106.231101
[arXiv:1011.1232 [hep-th]].

\bibitem{Hassan:2011hr}
S.~F.~Hassan and R.~A.~Rosen,
``Resolving the Ghost Problem in non-Linear Massive Gravity,''
Phys. Rev. Lett. \textbf{108}, 041101 (2012)
doi:10.1103/PhysRevLett.108.041101
[arXiv:1106.3344 [hep-th]].

\bibitem{deRham:2011qq}
C.~de Rham, G.~Gabadadze and A.~J.~Tolley,
``Helicity decomposition of ghost-free massive gravity,''
JHEP \textbf{11}, 093 (2011)
doi:10.1007/JHEP11(2011)093
[arXiv:1108.4521 [hep-th]].

\bibitem{Hassan:2011tf}
S.~F.~Hassan, R.~A.~Rosen and A.~Schmidt-May,
``Ghost-free Massive Gravity with a General Reference Metric,''
JHEP \textbf{02}, 026 (2012)
doi:10.1007/JHEP02(2012)026
[arXiv:1109.3230 [hep-th]].

\bibitem{Hassan:2011ea}
S.~F.~Hassan and R.~A.~Rosen,
``Confirmation of the Secondary Constraint and Absence of Ghost in Massive Gravity and Bimetric Gravity,''
JHEP \textbf{04}, 123 (2012)
doi:10.1007/JHEP04(2012)123
[arXiv:1111.2070 [hep-th]].

\bibitem{Mirbabayi:2011aa}
M.~Mirbabayi,
``A Proof Of Ghost Freedom In de Rham-Gabadadze-Tolley Massive Gravity,''
Phys. Rev. D \textbf{86}, 084006 (2012)
doi:10.1103/PhysRevD.86.084006
[arXiv:1112.1435 [hep-th]].

\bibitem{Golovnev:2011aa}
A.~Golovnev,
``On the Hamiltonian analysis of non-linear massive gravity,''
Phys. Lett. B \textbf{707}, 404-408 (2012)
doi:10.1016/j.physletb.2011.12.064
[arXiv:1112.2134 [gr-qc]].

\bibitem{Hassan:2012qv}
S.~F.~Hassan, A.~Schmidt-May and M.~von Strauss,
``Proof of Consistency of Nonlinear Massive Gravity in the St\"uckelberg Formulation,''
Phys. Lett. B \textbf{715}, 335-339 (2012)
doi:10.1016/j.physletb.2012.07.018
[arXiv:1203.5283 [hep-th]].

\bibitem{Deffayet:2012nr}
C.~Deffayet, J.~Mourad and G.~Zahariade,
``Covariant constraints in ghost free massive gravity,''
JCAP \textbf{01}, 032 (2013)
doi:10.1088/1475-7516/2013/01/032
[arXiv:1207.6338 [hep-th]].

\bibitem{ErrastiDiez:2019ttn}
V.~Errasti D\'iez, B.~Gording, J.~A.~M\'endez-Zavaleta and A.~Schmidt-May,
``Complete theory of Maxwell and Proca fields,''
Phys. Rev. D \textbf{101}, no.4, 045008 (2020)
doi:10.1103/PhysRevD.101.045008
[arXiv:1905.06967 [hep-th]].

\bibitem{ErrastiDiez:2019trb}
V.~Errasti D\'\i{}ez, B.~Gording, J.~A.~M\'endez-Zavaleta and A.~Schmidt-May,
``Maxwell-Proca theory: Definition and construction,''
Phys. Rev. D \textbf{101}, no.4, 045009 (2020)
doi:10.1103/PhysRevD.101.045009
[arXiv:1905.06968 [hep-th]].

\bibitem{ErrastiDiez:2021ykk}
V.~Errasti D\'\i{}ez and M.~K.~Marinkovic,
``Symplectic quantization of multifield generalized Proca electrodynamics,''
Phys. Rev. D \textbf{105}, no.10, 105022 (2022)
doi:10.1103/PhysRevD.105.105022
[arXiv:2112.11477 [hep-th]].

\bibitem{tHooft:1974pnl}
G.~'t Hooft,
``A Two-Dimensional Model for Mesons,''
Nucl. Phys. B \textbf{75}, 461-470 (1974)
doi:10.1016/0550-3213(74)90088-1.

\bibitem{Witten:1978ka}
E.~Witten,
``$\theta$ Vacua in Two-dimensional Quantum Chromodynamics,''
Nuovo Cim. A \textbf{51}, 325 (1979)
doi:10.1007/BF02776593.

\bibitem{Lemos:1993hz}
J.~P.~S.~Lemos and P.~M.~S\'a,
``The Two-dimensional analog of general relativity,''
Class. Quant. Grav. \textbf{11}, L11-L14 (1994)
doi:10.1088/0264-9381/11/1/003
[arXiv:gr-qc/9310041 [gr-qc]].

\bibitem{Deser:1995cs}
S.~Deser,
``Inequivalence of first and second order formulations in D = 2 gravity models,''
Found. Phys. \textbf{26}, 617 (1996)
doi:10.1007/BF02058235
[arXiv:gr-qc/9512022 [gr-qc]].

\bibitem{Wheeler}
N.~Wheeler,
``Electrodynamics in 2-Dimensional Spacetime,'' (1997)
http://kirkmcd.princeton.edu/examples/EM/wheeler\_2dem\_97.pdf.

\bibitem{McDonald}
K.~T.~McDonald,
``Electrodynamics in 1 and 2 Spatial Dimensions,'' (2019)
http://kirkmcd.princeton.edu/examples/2dem.pdf.

\end{thebibliography}
\end{document}